\documentclass{moriond}

\usepackage{amsmath} 
\usepackage{amssymb}
\usepackage{amsfonts}
\usepackage{ifthen}
\newboolean{inbibliography}
\setboolean{inbibliography}{false} 
\newboolean{uprightparticles}
\setboolean{uprightparticles}{false} 
\usepackage{xspace}
\usepackage{upgreek}
\usepackage{hyperxmp}
\usepackage{hyperref}
\usepackage[superscript,nomove]{cite} 
\usepackage{mciteplus}
\newboolean{articletitles}
\setboolean{articletitles}{false} 

\def\thebaroffset{0.18em}
\newcommand{\offsetoverline}[2][\thebaroffset]{\kern #1\overline{\kern -#1 #2}}%
\def\CP{{\ensuremath{C\!P}}\xspace}
\newcommand{\decay}[2]{\mbox{\ensuremath{#1\!\to #2}}\xspace}
\def\PK      {\ensuremath{K}\xspace}
\def\kaon    {{\ensuremath{\PK}}\xspace}
\def\Kp      {{\ensuremath{\kaon^+}}\xspace}
\def\Km      {{\ensuremath{\kaon^-}}\xspace}

\def\Ppi         {\ensuremath{\pi}\xspace}
\def\pion   {{\ensuremath{\Ppi}}\xspace}
\def\pip    {{\ensuremath{\pion^+}}\xspace}
\def\pim    {{\ensuremath{\pion^-}}\xspace}

\def\Dz      {{\ensuremath{\D^0}}\xspace}
\def\Dbar    {{\kern 0.2em\overline{\kern -0.2em \PD}{}}\xspace}
\def\Dzb     {{\ensuremath{\Dbar{}^0}}\xspace}
\def\dkk        {\decay{\Dz}{\Km\Kp}}
\def\dpipi      {\decay{\Dz}{\pim\pip}}

\newcommand{\aunit}[1]{\ensuremath{\text{\,#1}}}

\newcommand{\tev}{\aunit{Te\kern -0.1em V}\xspace}
\def\PD      {\ensuremath{D}\xspace}
\def\D       {{\ensuremath{\PD}}\xspace}

\def\PB      {\ensuremath{B}\xspace}
\def\B       {{\ensuremath{\PB}}\xspace}
\def\Bbar    {{\ensuremath{\kern 0.18em\overline{\kern -0.18em \PB}{}}}\xspace}

\newcommand{\deltaACP}{\ensuremath{\Delta A_{\CP}}\xspace}
\newcommand{\DACP}{\deltaACP}

\def\Dstarp  {{\ensuremath{\D^{*+}}}\xspace}
\def\Dstarm  {{\ensuremath{\D^{*-}}}\xspace}
\newcommand{\Araw}{\ensuremath{A_{\rm raw}}\xspace}

\newcommand{\mevcc}{\ensuremath{\aunit{Me\kern -0.1em V\!/}c^2}\xspace}

\def\Pnu         {\ensuremath{\nu}\xspace}

\newcommand{\chisq}{\ensuremath{\chi^2}\xspace}

\def\Ps      {\ensuremath{s}\xspace}
\def\squark    {{\ensuremath{\Ps}}\xspace}
\def\Bs      {{\ensuremath{\B^0_\squark}}\xspace}
\def\Bpm     {{\ensuremath{\B^\pm}}\xspace}

\def\PLambda     {\ensuremath{\Lambda}\xspace}
\mathchardef\PLambda="7103

\def\Dsp     {{\ensuremath{\D^+_\squark}}\xspace}

\def\Dp     {{\ensuremath{\D^+}}\xspace}

\def\squarkbar {{\ensuremath{\overline \squark}}\xspace}

\def\neub       {{\ensuremath{\overline{\Pnu}}}\xspace}
\def\neumb      {{\ensuremath{\neub_\mu}}\xspace}

\def\Pphi        {\ensuremath{\phi}\xspace}                 
\newcommand{\phiz}{\ensuremath{\Pphi}\xspace}
\def\DporDsp {{\ensuremath{\D_{(\squark)}^+}}\xspace}
\def\DmorDsm {{\ensuremath{\D{}_{(\squark)}^-}}\xspace}
\def\ACPKK {{\ensuremath{A_\CP(\Dz \to \Km \Kp)}}\xspace}
\def\ACPpipi {{\ensuremath{A_\CP(\Dz \to \pim \pip)}}\xspace}
\def\ACPhh {{\ensuremath{A_\CP(\Dz \to h^- h^+)}}\xspace}
\def\Kbar    {{\ensuremath{\offsetoverline{\PK}}}\xspace}
\def\Kbz     {{\ensuremath{\Kbar{}^0}}\xspace}

\def\phissss   {{\ensuremath{\phi_\squark^{\squark\squarkbar\squark}}}\xspace}

\newcommand{\Photo}{}

\begin{document}
\vspace*{4cm}
\title{CKM AND \boldmath{\CP} VIOLATION IN BEAUTY AND CHARM DECAYS IN LHCb}

\author{ F. BETTI on behalf of the LHCb collaboration}

\address{School of Physics and Astronomy, University of Edinburgh,\\
James Clerk Maxwell Building, Peter Guthrie Tait Road,\\
Edinburgh, EH9 3FD, United Kingdom}

\maketitle\abstracts{
    Measurements of \CP violation and Cabibbo-Kobayashi-Maskawa matrix in beauty and charm hadron decays are the core business of the LHCb physics programme.
    In this contribution, the most recent measurements performed by the LHCb collaboration on this topic are reported.
    The most precise measurement of time-dependent \CP asymmetry parameters in \decay{\Bs}{\phi\phi} decays has been done, providing results fully compatible with Standard Model expectations.
    The polarisation-dependent \CP-violation parameters of the same decay are measured for the first time.
    The combination of beauty and charm results gives $\gamma = \left( 63.8^{+3.5}_{-3.7} \right)^\circ$.
    The first search for local \CP violation in \decay{\DporDsp}{\Km \Kp \Kp} has been performed, resulting in no local \CP violation observed.
    The measurement of time-integrated \CP violation in \dkk decays, combined with previous measurements performed by LHCb, provides evidence of direct \CP violation in charm in a single decay channel at the level of $3.8$ standard deviations.
}

\section{Introduction}

In the Standard Model (SM) of particle physics, the combined symmetry of charge conjugation ($C$) and parity ($P$), collectively known as \CP, is violated by the weak interaction.
The \CP violation is one of the Sakharov conditions necessary for early universe baryogenesis, which, when satisfied, ensure a universe dominated by matter.~\cite{Sakharov:1967dj}
However, the amount of \CP violation measured experimentally so far is not enough to explain the matter-antimatter asymmetry observed in the universe.
Hence, the search for sources of \CP violation beyond the SM is strongly motivated, as well as the precise measurement of \CP violation within the SM itself.

Within the SM, \CP violation is quantified by the Cabibbo-Kobayashi-Maskawa (CKM) matrix,~\cite{Cabibbo:1963yz,Kobayashi:1973fv} a 3x3 unitary matrix that describes the weak interaction between up-type and down-type quarks.
The CKM matrix elements represent the transition amplitudes for quarks to change flavour, and must be determined through experimental observations.
The unitarity of the CKM matrix is a fundamental property within the SM framework.
The CKM matrix can be parameterised using three mixing angles and a single complex phase, where this phase accounts for all \CP violation in the quark sector within the SM.
Therefore, in order to accurately assess \CP violation in the SM, it is crucial to precisely measure the elements of the CKM matrix.
Moreover, experimental verification of the unitarity relations provides important tests of the SM and can reveal potential deviations from the SM expectations.

In this contribution, recent measurements performed by the LHCb collaboration concerning \CP violation in \decay{\Bs}{\phiz \phiz} decay and mixing, determination of the $\gamma$ angle and \CP violation in charm mesons are presented.

\section{\CP Violation in \decay{\Bs}{\phiz \phiz} decays}

\begin{figure}[t]
    \centering
    \includegraphics[width=0.45\textwidth]{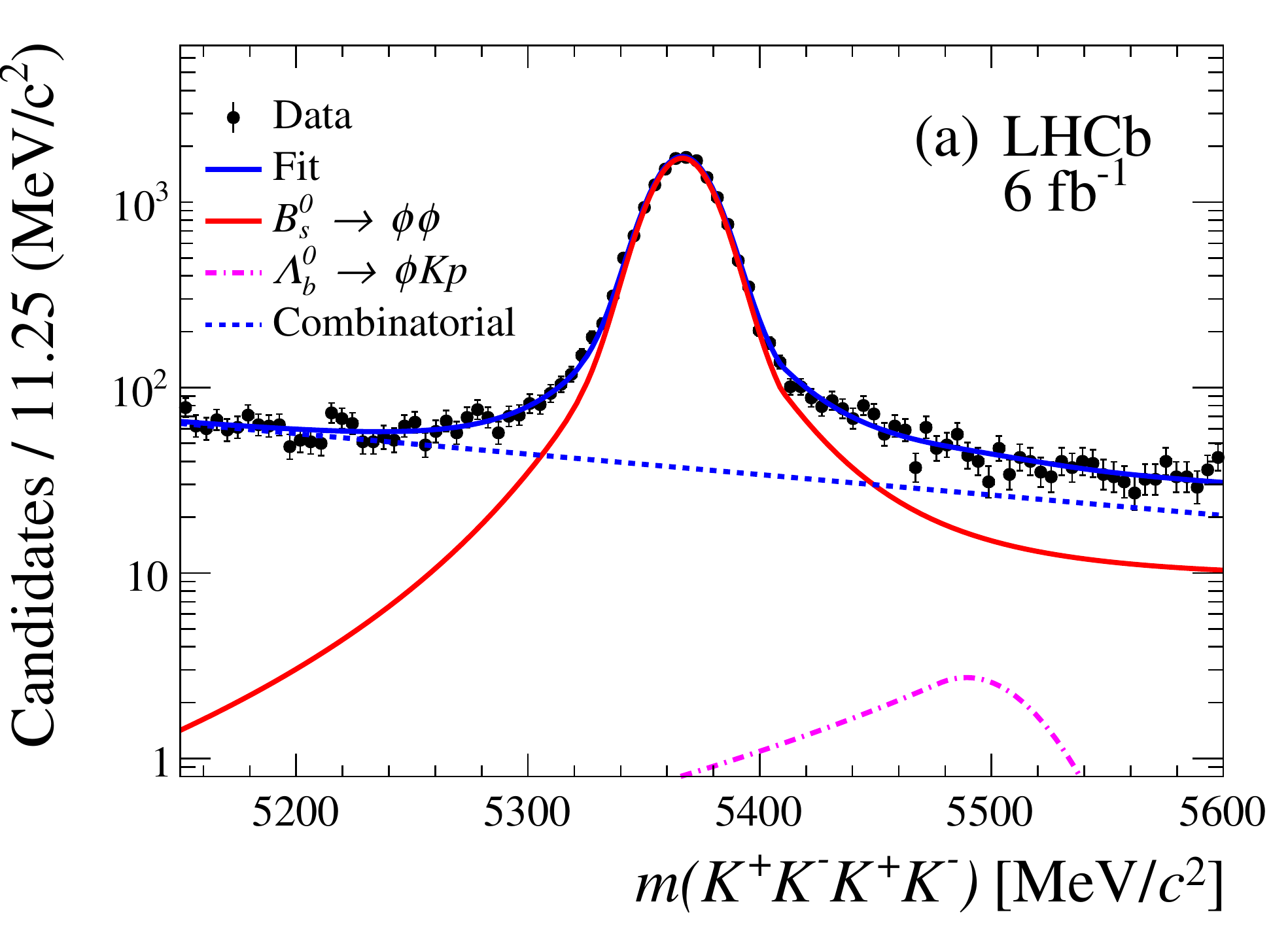}
    \includegraphics[width=0.45\textwidth]{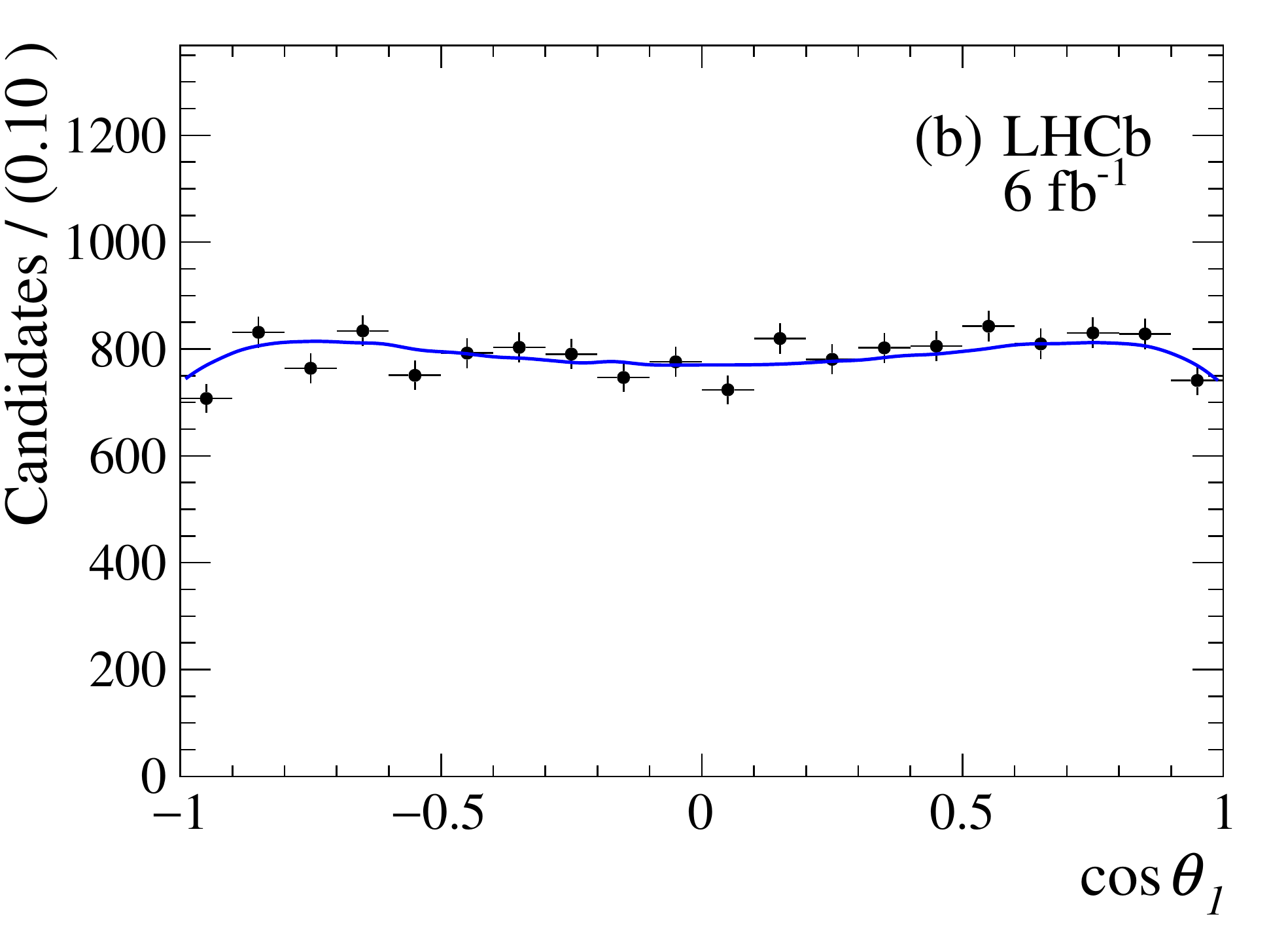}
    \includegraphics[width=0.45\textwidth]{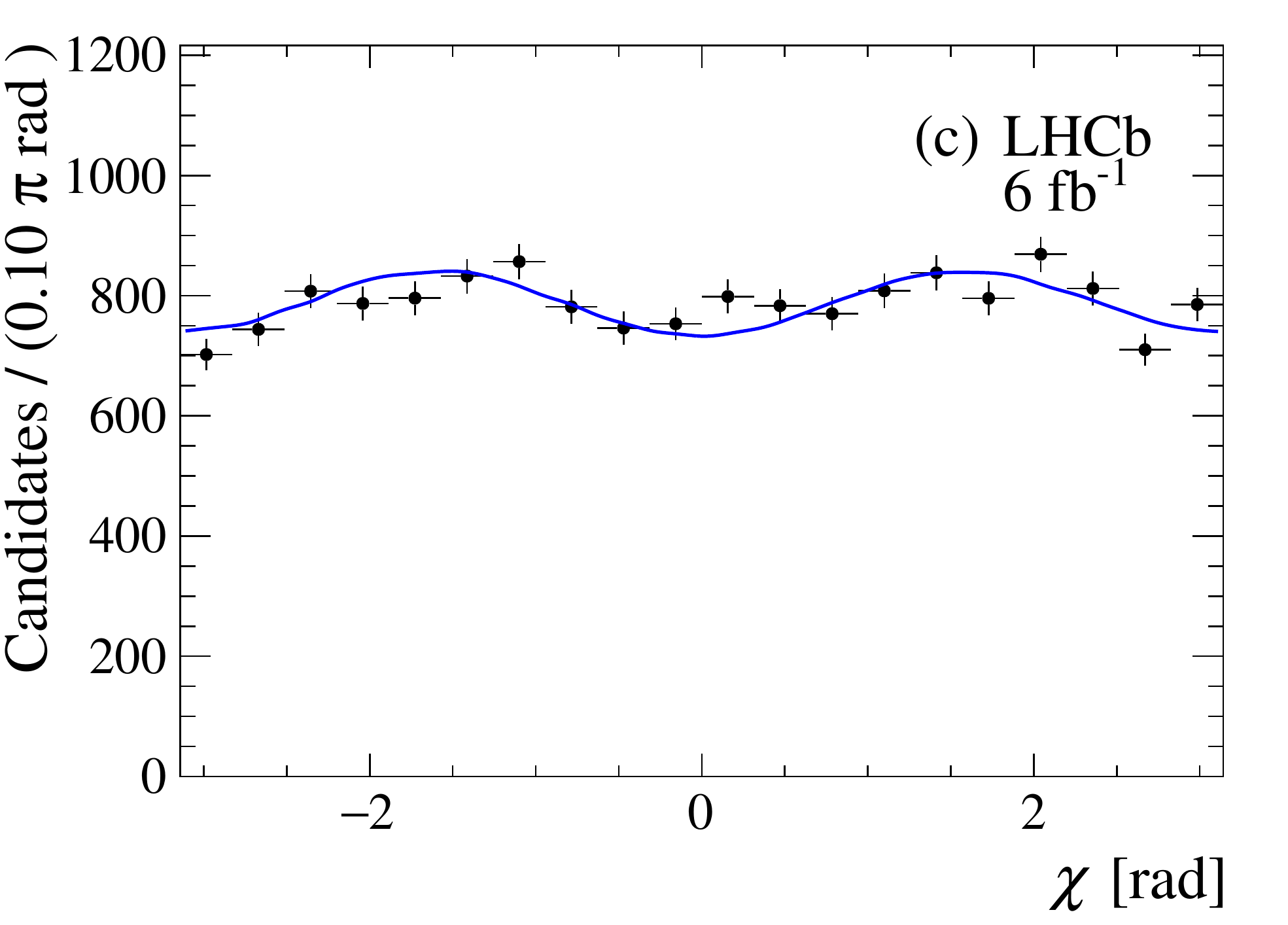}
    \includegraphics[width=0.45\textwidth]{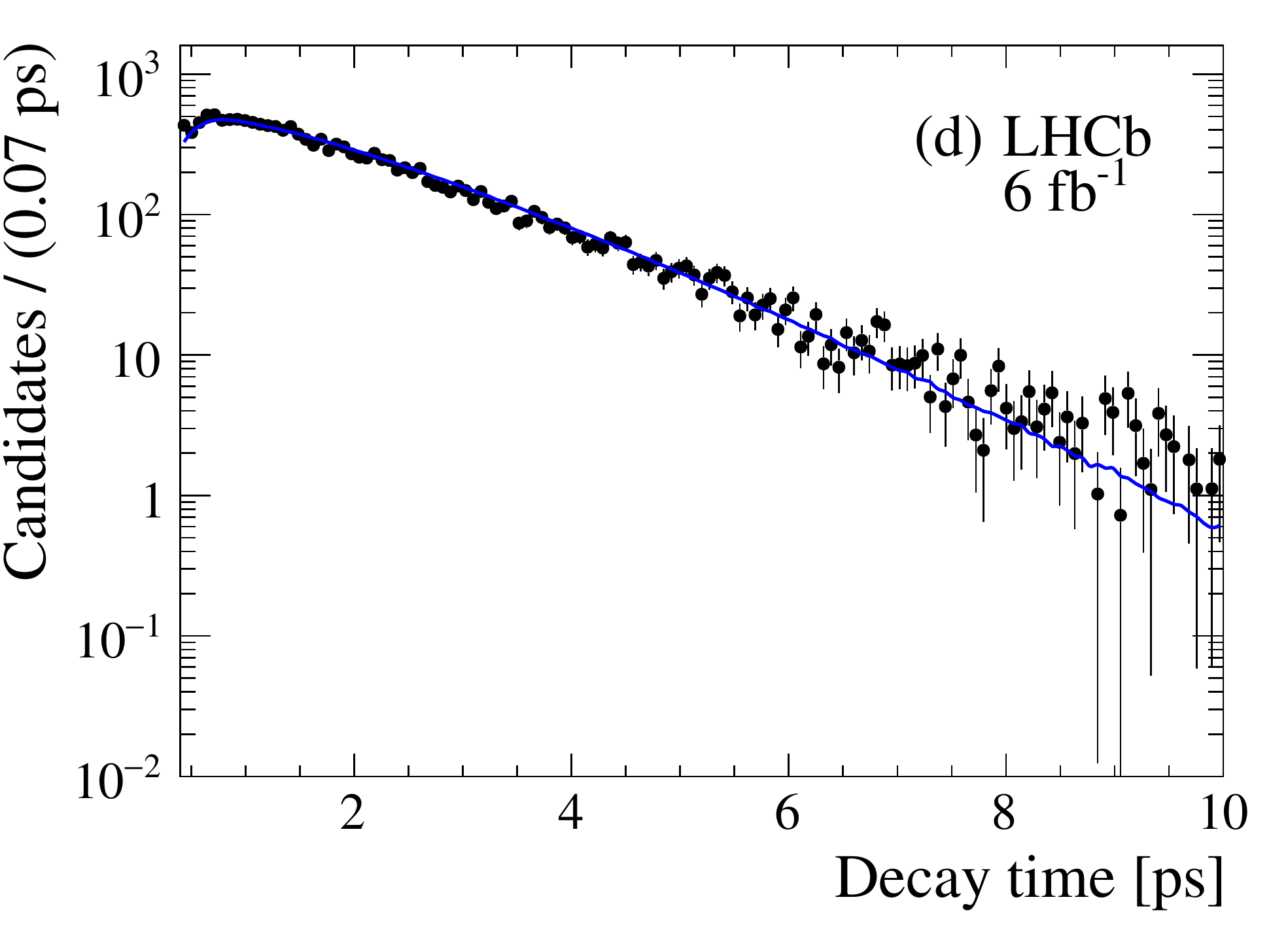}
    \caption{Mass distribution of the \decay{\Bs}{\phiz \phiz} candidates, superimposed by the fit projections (top left). Background-subtracted distributions of angular variables $\cos\theta_1$ (top right) and $\chi$ (bottom left), and decay time (bottom right), superimposed by the fit projections.}
    \label{fig:fit}
\end{figure}

In \decay{\Bs}{\phiz \phiz} decays, the time-dependent \CP violations arises from the interference between decay and mixing, characterised by the phase \phissss and the $|\lambda|$ parameter, that are expected to be very close to 0 and 1, respectively, in the SM.~\cite{Wang:2017rmh,Yan:2018fif}
New physics contributions in the penguin decay or the \Bs mixing could significantly alter the values of \phissss and $|\lambda|$.~\cite{Bhattacharya:2013sga,Kapoor:2023txh}
A new measurement carried out by the LHCb collaboration with the data collected during the Run~2 of the LHC has determined \phissss and $|\lambda|$.
The strategy consists in measuring the differential decay rate in terms of \Bs decay time and the three helicity angles that describe the three polarisation states of the final state.
The differential decay rate is a function of the polarisation-dependent parameters $\phi_{s,i}$ and $|\lambda_i|$ ($i = 0, \perp, \parallel$), so the measurement is able to search for new physics in different polarisation states.
The polarisation-independent measurement is performed by requiring $\phi_{s,i} = \phissss$ and $|\lambda_i| = |\lambda|$.
The signal candidates are statistically separated by the \decay{\Lambda_b^0}{\phiz K p} and the combinatorial backgrounds by means of a fit to the $\phiz \phiz$ invariant mass, shown in Figure~\ref{fig:fit}.
The background-subtracted multidimensional distribution of helicity angle and decay time is then fit by the differential decay rate.

The result of the polarisation-independent measurement is~\cite{LHCb-PAPER-2023-001}
\begin{align*}
    \phissss &= ( -0.042 \pm 0.075 \pm 0.009 )\ \mathrm{rad}, \\
    |\lambda| &= 1.004 \pm 0.030 \pm 0.009,
\end{align*}
where the first uncertainties are statistical and the second are systematic.
The combination with the Run~1 measurement~\cite{LHCb-PAPER-2014-026} gives
\begin{align*}
    \phissss &= ( -0.074 \pm 0.069 )\ \mathrm{rad}, \\
    |\lambda| &= 1.009 \pm 0.030,
\end{align*}
that is in agreement with the SM expectations.
This is the most precise measurement of time-dependent \CP asymmetry parameters in \decay{\Bs}{\phi\phi} decays up to date.
The polarisation-dependent \CP-violation parameters of this decay are measured for the first time:
\begin{align*}
    |\lambda_0| &= 1.02 \pm 0.17, \\
    |\lambda_\perp / \lambda_0| &= 0.97 \pm 0.22, \\
    |\lambda_\parallel / \lambda_0| &= 0.78 \pm 0.21, \\
    \phi_{s,0} &= ( -0.18 \pm 0.09 )\ \mathrm{rad}, \\
    \phi_{s,\parallel} - \phi_{s,0} &= ( 0.12 \pm 0.09 )\ \mathrm{rad}, \\
    \phi_{s,\perp} - \phi_{s,0} &= ( 0.17 \pm 0.09 )\ \mathrm{rad},
\end{align*}
resulting in no difference observed between different polarisation states.

\section{Direct measurements of $\gamma$}

The angle $\gamma$ of the CKM matrix, defined as
\begin{equation}
    \gamma \equiv \arg{\left( - \frac{V_{ud} V_{ub}^*}{V_{cd} V_{cb}^*} \right)},
\label{eq:gamma}
\end{equation}
can be measured in tree-level decays sensitive to the interference between $b \to c W$ and $b \to u W$ transition amplitudes.
In fact, $\gamma$ is determined by measuring asymmetries and yield ratios in \decay{\Bpm}{D (\to f_D) h^\pm} decays ($h^\pm = \pi^\pm, K^\pm$), where $f_D$ is a final state common to \Dz and \Dzb.

In multibody decays, the sensitivity is reduced by the differences between the favoured and suppressed decays in the structure of the intermediate resonances.
This effect is parameterised by the so-called coherence factor, that allows the maximum sensitivity to be reached when it is close to 1, whereas it totally dilutes the sensitivity when it is closed to 0.
In \decay{B^\pm}{D(\to K^\mp \pi^\pm \pi^\pm \pi^\mp) h^\pm} decays, performing the measurement in 4 bins of \D phase space allows a significant improvement on the sensitivity to $\gamma$ to be reached,~\cite{Evans:2019wza} because in some bins the coherence factor can be larger than the integrated one.

The measurement of $\gamma$ with \decay{B^\pm}{D(\to K^\mp \pi^\pm \pi^\pm \pi^\mp) h^\pm} decays performed by the LHCb collaboration with the full dataset gives:~\cite{LHCb-PAPER-2022-017}
\begin{equation*}
    \gamma = \left( 54.8^{+6.0+0.6+6.7}_{-5.8-0.6-4.3} \right)^\circ,
\end{equation*}
where the first uncertainty is statistical, the second is systematic and the third is due to the external inputs of \Dz mixing and hadronic decay parameters.~\cite{BESIII:2021eud,LHCb-PAPER-2015-057}
This is the second most precise determination of $\gamma$ from a single \D decay mode.
A large improvement on the uncertainty related to the external inputs is expected by including future LHCb determinations of \Dz mixing parameters and future measurements of hadronic \D-decay parameters performed with incoming BESIII data.

Since \decay{\Bpm}{D(\to K^\pm \pi^\mp) h^\pm} are sensitive also to the \Dz mixing and hadronic decay parameters, the LHCb collaboration started recently to perform the combination of all the measurements sensitive to $\gamma$ and charm parameters.
The most recent $\gamma$-charm combination of LHCb measurements has been carried out including the latest determinations of $\gamma$ with the \decay{B^\pm}{D(\to K^\mp \pi^\pm \pi^\pm \pi^\mp) h^\pm} and \decay{B^\pm}{D(\to h^\pm h'^\mp \pi^0) h^\pm} decay modes,~\cite{LHCb-PAPER-2022-017,LHCb-PAPER-2021-036} the measurement of \Dz mixing parameters and \CP violation parameters in mixing using \decay{\Dz}{h^- h^+}~\cite{LHCb-PAPER-2021-041} and $\Bbar \to \Dz(\to K_s^0 \pip \pim) \mu^- \neumb X$ decays,~\cite{LHCb-PAPER-2022-020} and the measurement of time-integrated \CP asymmetry in \dkk decays~\cite{LHCb-PAPER-2022-024} with the data collected during the Run~2 of the LHC.
The resulting value of $\gamma$ is~\cite{LHCb-CONF-2022-003}
\begin{equation*}
    \gamma = \left( 63.8^{+3.5}_{-3.7} \right)^\circ,
\end{equation*}
fully compatible with indirect determinations.~\cite{UTfit:2006vpt,Charles:2015gya}.
The \Dz mixing parameters are determined to be
\begin{equation*}
    x = (0.398^{+0.050}_{-0.049})\%, \quad y = (0.636^{+0.020}_{-0.019})\%.
\end{equation*}

The latest determination of $\gamma$, not included in the $\gamma$-charm combination, has been done by the LHCb collaboration by using the $B^\pm \to D(\to \Kp \Km \pi^\pm \pi^\mp)h^\pm$ and \decay{B^\pm}{D(\to \pip \pim \pi^\pm \pi^\mp)h^\pm} decay modes with the data collected during the Run~1 and Run~2 of the LHC.
A phase-space integrated measurement is done in both decay channels, while a binned analysis is performed for the \decay{D}{\Kp \Km \pi^\pm \pi^\mp} mode, resulting in~\cite{LHCb-PAPER-2022-037}
\begin{equation*}
    \gamma = \left( 116^{+12}_{-14} \right)^\circ.
\end{equation*}
The precision of this measurement will be improved once new charm model-independent measurements of the \Dz decay parameters will be available.

\section{\CP violation in charm decays}

Charm decay are an unique laboratory to study \CP violation in up-type quark decays.
Due to the smallness of involved CKM elements and GIM mechanism, \CP violation in charm decays is predicted to be small.
Moreover, the long distance contributions are not negligible in the corresponding SM calculations, making very difficult to provide precise expectations for the related observables.
The measured value of the difference between the time-integrated \CP asymmetries in \dkk and \dpipi decays by the LHCb collaboration~\cite{LHCb-PAPER-2019-006} challenges the calculations based on first-principles QCD.
Further measurements are therefore needed in charm sector in order to clarify the picture.

Multi-body decays are particularly interesting, because phase-space local \CP asymmetries can be larger than integrated ones.
The LHCb collaboration has recently performed the first search for local \CP violation in \decay{\DporDsp}{\Km \Kp \Kp} decays using the data collected between 2016 and 2018.
The analysis makes use of the \textit{Miranda} technique,~\cite{BaBar:2008xzl,Bediaga:2009tr} that consists in dividing the Dalitz plot in bins and computing, for each bin $i$, the significance $S_\CP^i$ of the difference in the numbers of $\DporDsp$ candidates $N^i(\DporDsp)$ and $\DmorDsm$ candidates $N^i(\DmorDsm)$, where the latter is corrected for global charge asymmetry $\alpha \equiv \sum_i N^i(\DporDsp) / \sum_i N^i(\DmorDsm)$:
\begin{equation}
    S_\CP^i \equiv \frac{N^i(\DporDsp) - \alpha \cdot N^i(\DmorDsm)}{\sqrt{\alpha \left( \delta^2_{N^i(\DporDsp)} + \delta^2_{N^i(\DmorDsm)} \right)}},
\label{eq:SiCP}
\end{equation}
where $\delta_{N^i(\DporDsp)}$ and $\delta_{N^i(\DmorDsm)}$ are the statistical uncertainties of the signal candidate yields.
The yields are obtained by fitting the \DporDsp invariant mass distributions in each bin.
In the absence of local \CP violation, the values of $S_\CP^i$ follow a normal distribution.
A \chisq test is performed, with $\chisq \equiv \sum_i (S_\CP^i)^2$, and the corresponding $p$-value is calculated as the probability of obtaining a \chisq that is at least as high as the value observed, under the assumption of \CP conservation (null hypothesis).
If $p$-values are less than $3\times 10^{-7}$, \CP violation is observed.
The analysis, validated on Cabibbo-favoured \decay{\Dp}{\Km \pip \pip} and \decay{\Dsp}{\Km \Kp \pip} control modes, where no direct \CP violation is expected, results in a $p$-value of $13.3\%$ for the \Dsp mode and $31.6\%$ for the \Dp mode, corresponding to no evidence of local \CP violation.~\cite{LHCb-PAPER-2022-042}

The most recent measurement of \CP violation in two-body charm decays has been carried out by the LHCb collaboration with the data collected during the Run~2 of the LHC.
The flavor of the \Dz meson at the production is obtained by looking at the charge of the pion in the \decay{\Dstarp}{\Dz \pip} decay.
Since the raw asymmetry
\begin{equation}
    \Araw(\dkk) \equiv \frac {N\left(\Dstarp \to \Dz(\to \Km \Kp)\pip \right) - N\left(\Dstarm \to \Dzb(\to \Km \Kp)\pim\right)} {N\left(\Dstarp \to \Dz(\to \Km \Kp)\pip \right) + N\left(\Dstarm \to \Dzb(\to \Km \Kp)\pim \right)}
\label{eq:Araw}
\end{equation}
is affected by the detection asymmetry of the tagging pion and the production asymmetry of the \Dz meson, Cabibbo-favoured control modes are used to cancel the nuisance asymmetries and obtain the physical \CP asymmetry \ACPKK.
Two almost statistically independent calibration procedures are used: one, already used in the Run~1 measurements,~\cite{LHCb-PAPER-2014-013,LHCb-PAPER-2016-035} makes use of \decay{\Dp}{\Km \pip \pip} and \decay{\Dp}{\Kbz \pip} decays, whereas the other, which is used for the first time, exploits the \decay{\Dsp}{\phi \pip} and \decay{\Dsp}{\Kbz \Kp} decays.
The \decay{\Dstarp}{\Dz(\to \Km \pip) \pip} decay is needed by both calibration procedures.
The resulting time-integrated \CP asymmetries are~\cite{LHCb-PAPER-2022-024}
\begin{align*}
    \ACPKK | \Dp &= (13.6 \pm 8.8 \pm 1.6) \times 10^{-4}, \\
    \ACPKK | \Dsp &= (2.8 \pm 6.7 \pm 2.0) \times 10^{-4},
\end{align*}
for the first and second calibration procedure respectively, where the first uncertainties are statistical and the second are systematic, with statistical (systematic) correlation equal to 0.05 (0.28).
The time-integrated \CP asymmetry in the \decay{\Dz}{h^- h^+} decay is equal to
\begin{equation}
    \ACPhh = a_{h^- h^+}^\mathrm{d} + \frac{\langle t \rangle_{h^- h^+}}{\tau_\Dz} \Delta Y,
\label{eq:ACPhh}
\end{equation}
where $a_{h^- h^+}^\mathrm{d}$ is the direct \CP asymmetry in the decay, $\Delta Y$ parameterises the time-dependent \CP asymmetry, $\langle t \rangle_{h^- h^+}$ is the observed average \Dz decay time and $\tau_\Dz$ is the nominal \Dz lifetime.~\cite{PDG2022}
By combining all the LHCb measurements of \ACPKK, $\DACP \equiv \ACPKK - \ACPpipi$, $\Delta Y$ and $\langle t \rangle_{h^- h^+}$, the direct \CP asymmetries are obtained:
\begin{align*}
    a_{\Km \Kp}^\mathrm{d} &= (7.7 \pm 5.7) \times 10^{-4}, \\
    a_{\pim \pip}^\mathrm{d} &= (23.2 \pm 6.1) \times 10^{-4}, 
\end{align*}
with a correlation of 0.88.
This result corresponds to an evidence of direct \CP violation in \decay{\Dz}{\pim \pip} decays at the level of $3.8$ standard deviations, and exceeds the SM expectations of $U$-spin symmetry breaking at the level of 2 standard deviations.

\section{Conclusions}

LHCb continues to produce world-leading results on \CP violation in beauty and charm decays.
A new precise test of the SM in \Bs mixing has been made.
The CKM angle $\gamma$ is now known with an uncertainty smaller than $4^\circ$.
The first evidence of \CP violation in charm in a single decay channel at the level of $3.8$ standard deviations has been reported, and a new search of local \CP violation in charm hadron multi-body decay has been performed.
The LHCb Upgrade I will improve such measurements with the data collected during the Run~3 of the LHC, thanks to the higher integrated luminosity and a better trigger efficiency.

\section*{References}

\setboolean{inbibliography}{true}
\bibliography{bettifederico}

\end{document}